\documentclass[jkps,nofootinbib,twocolumn,fleqn,showkeys]{revtex4}
\usepackage{graphicx}
\usepackage{amssymb}
\usepackage{amsmath}
\usepackage{bm}
\usepackage{url}

\begin{document}
\setcounter{page}{0}
\title[]{Refinement for community structures of bipartite networks}
\author{Sang Hoon \surname{Lee}}
\email{lshlj82@gnu.ac.kr}
%\thanks{Fax: +82-2-554-1643}
\affiliation{Department of Physics and Research Institute of Natural Science, Gyeongsang National University, Jinju 52828, Korea}
\affiliation{Future Convergence Technology Research Institute, Gyeongsang National University, Jinju 52849, Korea}

%\date[]{Received 6 January 2021}
\date{\today}

\begin{abstract}
Bipartite networks composed of dichotomous node sets are ubiquitous in nature and society. Partly for simplicity's sake, many studies have focused on their projection onto their unipartite versions where one only needs to care about a single type of node. When it comes to mesoscale structures such as communities, however, properly incorporating \emph{a priori} structural restrictions such as bipartivity is ever more important. In this paper, as a case study, we take the community structure of bipartite networks in various scales to examine the amount of information of bipartivity encoded in the community detection procedure. In particular, we report the robustness in reliability of detected community based on consistency by comparing the detection algorithm with or without the consideration of bipartivity. From the analysis with model networks embedding prescribed communities and real networks, we find that the community detection tailored to take the bipartivity into account clearly yields more robust community structures than the one without such structural information. This demonstrates the necessity for customizing the community detection algorithm by encoding whatever information is known about networks of interest and, at the same time, raises an interesting question on the possibility of estimating the quantitative amount of information from such a customization.
\end{abstract}


\maketitle

\section{\label{sec:intro}INTRODUCTION}

Network science\cite{NewmanBook} has become an essential framework to investigate interacting ``complex''\footnote{The author is, in fact, reluctant to use this buzzword at this point, although everyone seems to acknowledge the existence of such systems without rigorous mathematical definition, including the Nobel Prize in Physics 2021\cite{NobelPrize}.} systems. Physicists naturally tend to seek the most universal features in such systems, so it is no surprise that the journey toward real interacting systems began from the most abstract mathematical entity described by graph theory\cite{GraphTheory} in mathematics. On the way to reality, however, they have encountered various types of constraints imposed in different systems. For instance, a network may represent relations in a physical space where one must take spatial constraints into account\cite{YLiu2020}, e.g., spatial networks\cite{SpaitlaNetworkReview} with physical restrictions such as the no edge-crossing rule\cite{SHLee2013}. One of the most prominent types of networks with such restrictions is the bipartite network\cite{NewmanBook,Holme2003}: a network composed of two distinct groups of nodes, where edges can only connect nodes belonging to different groups. Despite their seemingly artificial mathematical construction, bipartite networks describe many real interacting systems in nature and society, as their structure is applicable to any type of ``acts'' and their ``actors.'' Those acts and actors represent fundamental causal relations, and the concept of networks is embedded in the fact that multiple actors are involved in multiple acts together. In that respect, one might even claim that the unipartite network is a special type of network in which the sets of acts and actors happen to be equivalent, in addition to a more conventional notion of projection (connecting actors involved in the same act and connecting acts played by the same actor) from the bipartite network\cite{NewmanBook}. Examples include (literally) actors and movies, customers and merchandise, proteins and protein complexes, predators and prey, pollinators and plants, just to name a few. One can check its prevalence at the Colorado Index of Complex Networks (ICON) database\cite{ICON}, where $146$ networks explicitly labeled as ``bipartite'' are found (on October 8, 2021). For comparison, there are $149$ ``biological'' networks of any type in the same database (accessed on the same day). 

Considering the prevalence of such bipartite structures (or at least with some degree of bipartivity constraint\cite{Holme2003}), mesoscale sub-structures such as the popular topic of community structures\cite{Porter2009,Fortunato2010} of bipartite networks have been relatively underemphasized. Some properties of unipartite networks could just be directly transferable to bipartite ones, e.g., the number of connected neighbors called the degree\footnote{Even in the case of degree, however, one has to distinguish between the degrees of one type of node from those of the other type, as their distributions and the resultant implications can be very different\cite{SHLee2011}.}. Mesoscale properties, however, usually take \emph{groups of multiple nodes}, and grouping nodes with fundamentally different features together is of course nontrivial and requires a baseline assumption on which the criterion of clustering is based. A popular framework of identifying communities in networks relies on the notion of modularity maximization\cite{Porter2009,Fortunato2010,Newman2004}, where the edges inside trial communities literally count compared with the edges connecting different trial communities. Note that there seems to be no explicit conceptual ingredient related to uni- or bipartivity in this framework, but there is an important factor in the modularity function called the null-model term\cite{Bassett2013}, encoding the expected or ``background'' connectivity obviously determined by the presumed constraint in connectivity\cite{Bassett2013,SHLee2019} or lack thereof. Previous literature includes the explicit consideration of bipartivity\cite{Barber2007,Bartlett2015,Borge-Holthoefer2017} (the bipartite network of twitter users and hashtags in Ref.\cite{Borge-Holthoefer2017}), but many works just use the original null-model term for unipartite networks\cite{WCai2020}, including the author's own works\footnote{For the record, the present work includes the result of re-analyzing the data used in Ref.\cite{SHLee2020} and its implication.}\cite{SHLee2010,SHLee2020}.

In this work, we precisely take the issue of the null-model term modified to detect communities properly in the bipartivity networks introduced in Ref.\cite{Barber2007}, where the author suggested an appropriate functional form of null-model term for bipartite networks, but there was no comparative performance analysis between the original null-model term and the modified one. A rather tricky part here is in fact ``properly,'' as there is no unanimous test to evaluate quantitatively the superiority or the inferiority of detected network communities. For that purpose, we utilize a recently developed tool measuring the self-(in)consistency of network communities\cite{HKim2019,DLee2021} and present, for the model networks with different compositions of planted ground-truth communities (another missing thing in Ref.\cite{Barber2007}) followed by real networks, comparative robustness of the community detection based on the null-model term modified for bipartite networks\cite{Barber2007}. The result, compared with the original null-model term without the consideration of bipartite structures,  indicates a robustness of detected communities based on the properly tailor-made null-model term, which in turn emphasizes the importance of embedding information available for mesoscale network analysis. 

\section{\label{sec:theory}Theoretical framework}

\subsection{\label{sec:modified_null_model}Modified null-model term for bipartite networks}

\begin{figure*}
\includegraphics[width=0.75\textwidth]{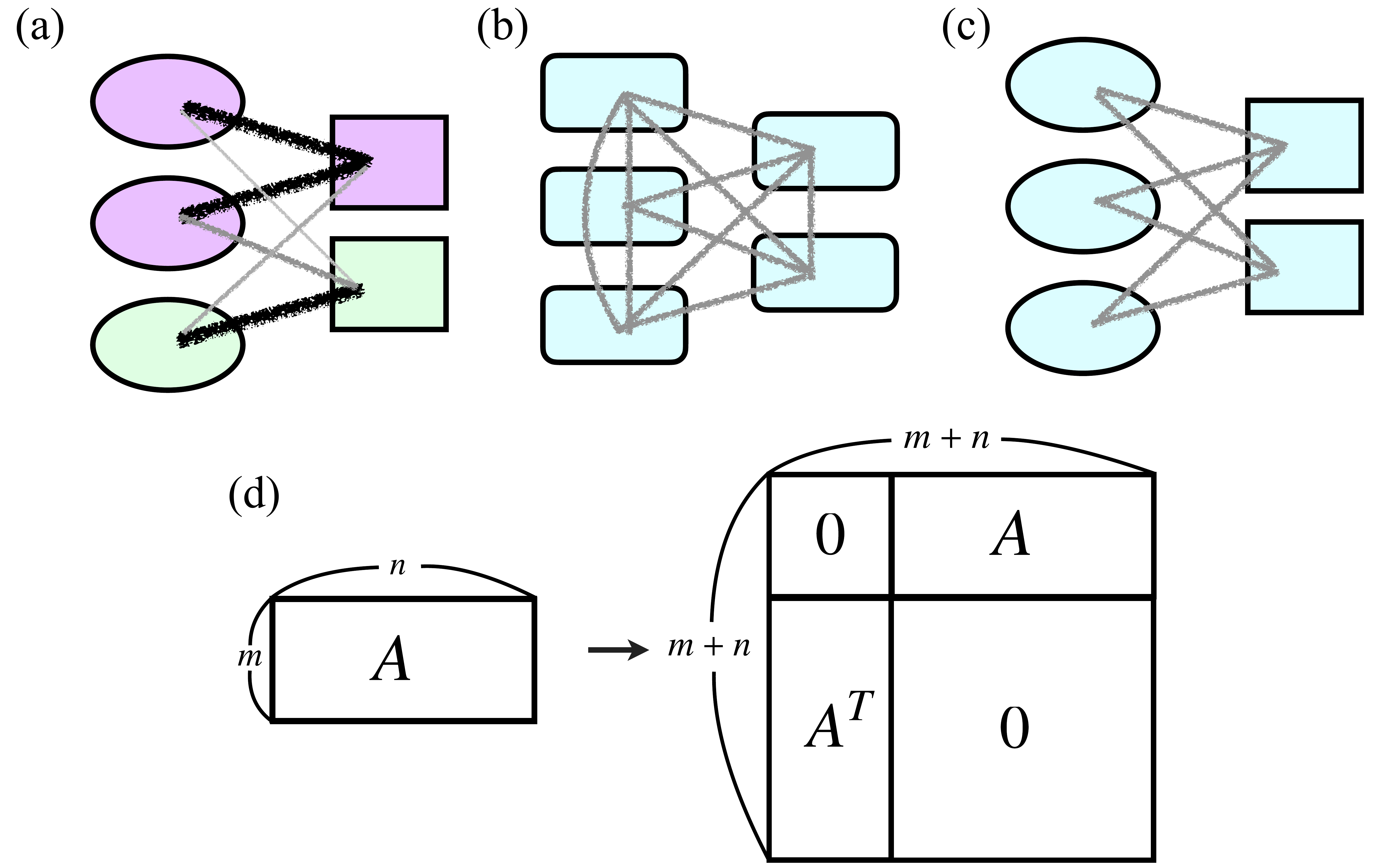} 
\caption{Illustration of (a) an example of modular structures (colored with light purple above and light green below) in a bipartite network with nodes of an oval type on the left and nodes of a rectangular type on the right, and the corresponding null models for (b) the NG type\cite{Newman2004} and (c) the BB type\cite{Barber2007}. The thickness and shade indicate the amount of intermodular connection. In practice, we realize the BB-type null model by constructing the augmented square matrix (and its null-model matrix by replacing $A$ with its NG null-model counterpart while keeping the two empty squares denoted by ``0'') illustrated in the panel (d), as discussed in Ref.\cite{Barber2007}.} 
\label{fig:illustration}
\end{figure*}

Modularity-based network community detection\cite{Porter2009,Fortunato2010,Newman2004} is arguably the most popular framework for identifying communities in networks, partly because of its already established status quo including various numerical tools on top of its mathematically intuitive interpretation of detecting densely connected subsets of nodes. The most widely used type of modularity function for a given community partition $\{ g_i \}$ ($g_i$ represents the community identity of node $i$) is given by
\begin{equation}
Q = \frac{1}{2m} \sum_{i \ne j} \left[ \left( A_{ij} - \gamma \frac{k_i k_j}{2m} \right) \delta(g_i,g_j) \right ] \,,
\label{eq:modularity}
\end{equation}
where the adjacency matrix elements $A_{ij}$ represent the existence of an edge between nodes $i$ and $j$ ($A_{ij} = 1$ if the edge exists and $A_{ij} = 0$ if it does not)\footnote{We only consider undirected networks with $A_{ij} = A_{ji}$ for every node pairs $(i,j)$, as the concept of communities in directed networks can be quite nontrivial. For the interested readers, see Refs.\cite{Leicht2008,YKim2010}}, and the degree $k_i$ represents the number of neighbors of node $i$ and is given by $k_i = \sum_j A_{ij}$. The normalization constant is $\sum_i k_i = 2m$, and the summand is counted only when the Kronecker delta term $\delta(g_i, g_j)$ is turned on for the node pairs $i$ and $j$ belonging to the same community. In the case of weighted networks, the elements $A_{ij}$ represents the weight on the edge between nodes $i$ and $j$ ($A_{ij}=0$ when the edge between nodes $i$ and $j$ is absent), and the strength $k_i$ represents the sum of the weights around node $i$, i.e., $k_i = \sum_j A_{ij}$ (the same formula as the degree in the case of unweighted networks). An important assumption behind the modularity function in Eq.~\eqref{eq:modularity} is that the contribution of each edge to the modularity is subtracted by the term $k_i k_j / (2m)$ called the Newman-Girvan (NG) null-model term named after the authors of Ref.\cite{Newman2004} and corresponds to the expected probability of the edge's existence \emph{judged solely from each node's degree}. A practically useful parameter here is the resolution parameter $\gamma$ in front of the null-model term, which controls the overall scale of detected communities\cite{Porter2009,Fortunato2010,Reichardt2006,Onnela2012}. The role of $\gamma$ was first found empirically\cite{Reichardt2006}, but the parameter has been demonstrated to be indeed related to the parameters controlling the number of preassigned communities in the stochastic block model\cite{Newman2016}. 

As a reminder, the mathematical form of the NG null-model term $k_i k_j / (2m)$ in Eq.~\eqref{eq:modularity} comes from the assumption of our ignorance except for the degree information: the larger the degree of a node is, the more likely the node will be connected to \emph{any} other node. In reality, of course, one may get access to additional information that can sharpen the null-model term. Therefore, in general, the modularity function can be generalized as
\begin{equation}
Q = \frac{1}{2m} \sum_{i \ne j} \left[ \left( A_{ij} - \gamma P_{ij} \right) \delta(g_i,g_j) \right ] \,,
\label{eq:Gmodularity}
\end{equation}
where $P_{ij}$ is the generalized null-model term representing the expected probability of the existence of an edge connecting nodes $i$ and $j$ from all of the information available to us\cite{Bassett2013}. For instance, a network can be embedded in space, and one may know the functional form of the distance-dependent connection probability, which can be utilized to set $P_{ij}$ to encode the spatial information\cite{SHLee2019}. 

For bipartite networks composed of two distinct types of nodes, the structural constraint of not allowing edges between the nodes of the same type can directly be used to modify $P_{ij}$. The specific type of null-model for bipartite networks called the Barber (BB) type null-model \cite{Barber2007} (named after the author of Ref.\cite{Barber2007}, of course), used in this work is illustrated in Fig.~\ref{fig:illustration}. Suppose that we have a given community structure (the light purple community above and the light green community below) in the bipartite network in Fig.~\ref{fig:illustration}(a). The NG null-model term in Eq.~\eqref{eq:modularity} corresponds to the ``average'' connection illustrated in Fig.~\ref{fig:illustration}(b) regardless of node types, which obviously allows a connection between nodes of the same type that would never exist in reality by construction. In other words, if we just use the original modularity in Eq.~\eqref{eq:modularity} to detect communities in bipartite networks, the modularity function $Q$ will include meaningless negative values in the summand, which would require unnecessary computational time and may systematically produce a misleading result in the worse case. In contrast, if we use the BB type null-model term corresponding to Fig.~\ref{fig:illustration}(c), the expected edges only connect node pairs from different types of nodes, and the modified modularity function in Eq.~\eqref{eq:Gmodularity} with $P_{ij}$ in Ref.\cite{Barber2007} [$P_{ij} = 0$ for all of the node pairs $(i,j)$ from the same type and $P_{ij} = k_i k_j / (2m)$ if nodes $i$ and $j$ are of different types] will yield communities more efficiently and more accurately. In practice, as illustrated in Fig.~\ref{fig:illustration}(d), we can directly use various types of modularity maximization algorithms for detecting communities in unipartite networks by constructing the augmented square matrix and its NG null-model counterpart for each rectangular blocks (equivalent to the BB-type null-model term).

Among modularity-maximization-based community detection algorithms, we use the GenLouvain algorithm\cite{GenLouvain}, which is a variant of the popular Louvain algorithm\cite{Louvain}. The algorithm is known for its convenient implementation and computational efficiency, but more importantly in this work, it is a stochastic algorithm allowing in principle different detection results for each realization. How can such a property be beneficial at all? We will look into that now.

\subsection{\label{sec:inconsistency}Inconsistency analysis of communities}

One of the most tricky parts of network community detection is that in most cases, no ground-truth community division exists in the first place. In other words, finding communities is not searching for a rigorously defined mathematical object, but is close to trying an ensemble of divisions for a given network structure, the ultimate goal of which is the expectation of extracting practically useful results. In this spirit, throughout the series of works\cite{HKim2019,DLee2021} in which the author participated, we have demonstrated that multiple different results of community detection, usually expressed as community inconsistency\cite{HKwak2011,Lancichinetti2012}, are not necessarily a malfunction of stochastic algorithms that should be remedied. Local inspection of such an inconsistency provides valuable information on an individual node's function as bridges connecting communities\cite{HKim2019}, and more importantly in this work, global inspection of community inconsistency effectively plays the role of a statistical reliability test, especially for a range of community resolution adjusted by using the $\gamma$ parameter in Eqs.~\eqref{eq:modularity} or \eqref{eq:Gmodularity}. 

Of course, due to the fundamental limit of community detection mentioned in the first part of this subsection, we should take the reliability as a measure of self-consistency rather than a ground-truth solution, but it still provides practically useful evidence on the validity of detected communities. In this work, we take the global inconsistency measure called the partition inconsistency (PaI) curve $\Omega$ as a function of the resolution parameter $\gamma$ introduced in Ref.\cite{DLee2021} and utilize it as the main inspection tool for comparative analysis on the effect of different null-model terms for bipartite networks. 
The PaI quantifies how inconsistently communities are detected for a given ensemble of results from a stochastic network community detection algorithm. 

From the total number $m$ of community detection realizations, we denote the number of unique configurations detected by $\mathcal{C}$ and the proportion of each configuration $\alpha \in \{ 1, 2, \cdots, \mathcal{C} \}$ by $p_{\alpha} = m_{\alpha} / m$, where $m_{\alpha}$ is the number of configurations that include $\alpha$.
The similarity $S_{\alpha\beta}$ between configurations $\alpha$ and $\beta$ is defined by employing the element-centric similarity (ECS)\footnote{The ECS measure introduced in Ref.\cite{EC} has a number of advantages over other conventional measures of similarity between different community configurations such as normalized mutual information (NMI), so we strongly encourage the practitioners of network community detection to try different measures instead of sticking to a single measure just because it is widely used.}\cite{EC}. In the cluster-induced element graph of configuration $\alpha$, we assign an attribute of node $i$ with respect to another node $j$, which is the stationary probability distribution $f_{ij}^{\alpha}$ induced by the personalized PageRank algorithm (PPR)\cite{PPR} applied to node $i$. 
Then, the ECS between two community configurations $\alpha$ and $\beta$ is defined as
\begin{equation}\label{eq:EC}
S_{\alpha\beta} = \frac{1}{N}\sum_{i=1}^{N} \left( 1 - \frac{1}{2d}\sum_{j=1}^{N} \left| f_{ij}^{\alpha} - f_{ij}^{\beta}\right| \right) \,,
\end{equation}
where $N$ is the number of nodes and $f_{ij}^{\alpha}$ is node $j$'s relative importance in the stationary state of PPR starting from the node $i$ in configuration $\alpha$. We use the default value of the damping factor $d = 0.9$ as used in Ref.\cite{EC}.
Based on this similarity measure, we define the PaI of a community ensemble as
\begin{equation}\label{eq:PaI}
\Omega = \left( \sum_{\alpha=1}^{\mathcal{C}}\sum_{\beta=1}^{\mathcal{C}} p_{\alpha}p_{\beta}S_{\alpha \beta}\right)^{-1} \,,
\end{equation}
which corresponds to the reciprocal of the average similarity (weighted by the relative appearance $\{ p_\alpha \}$) between all configuration pairs. 
If all the configurations are independent of the others and emerge with the same uniform probability $p_{\alpha} = 1 / \mathcal{C}$, all of the off-diagonal components of $S_{\alpha \beta}$ vanish, and the PaI of this ensemble reaches the maximum value of $\mathcal{C}$. 
In contrast, if every realization of the community detection gives exactly the same community structure, it becomes trivially $\Omega = 1$ (the minimally inconsistent or the maximally consistent case). We will now see if the more sophisticated null-model term introduced by Barber\cite{Barber2007} indeed gives more reliable community divisions than the original NG null-model term\cite{Newman2004}.

\section{\label{sec:results}Results}

\subsection{\label{sec:model}Model networks}

\begin{figure*}
\begin{tabular}{ll}
(a) & (b) \\
\includegraphics[width=0.45\textwidth]{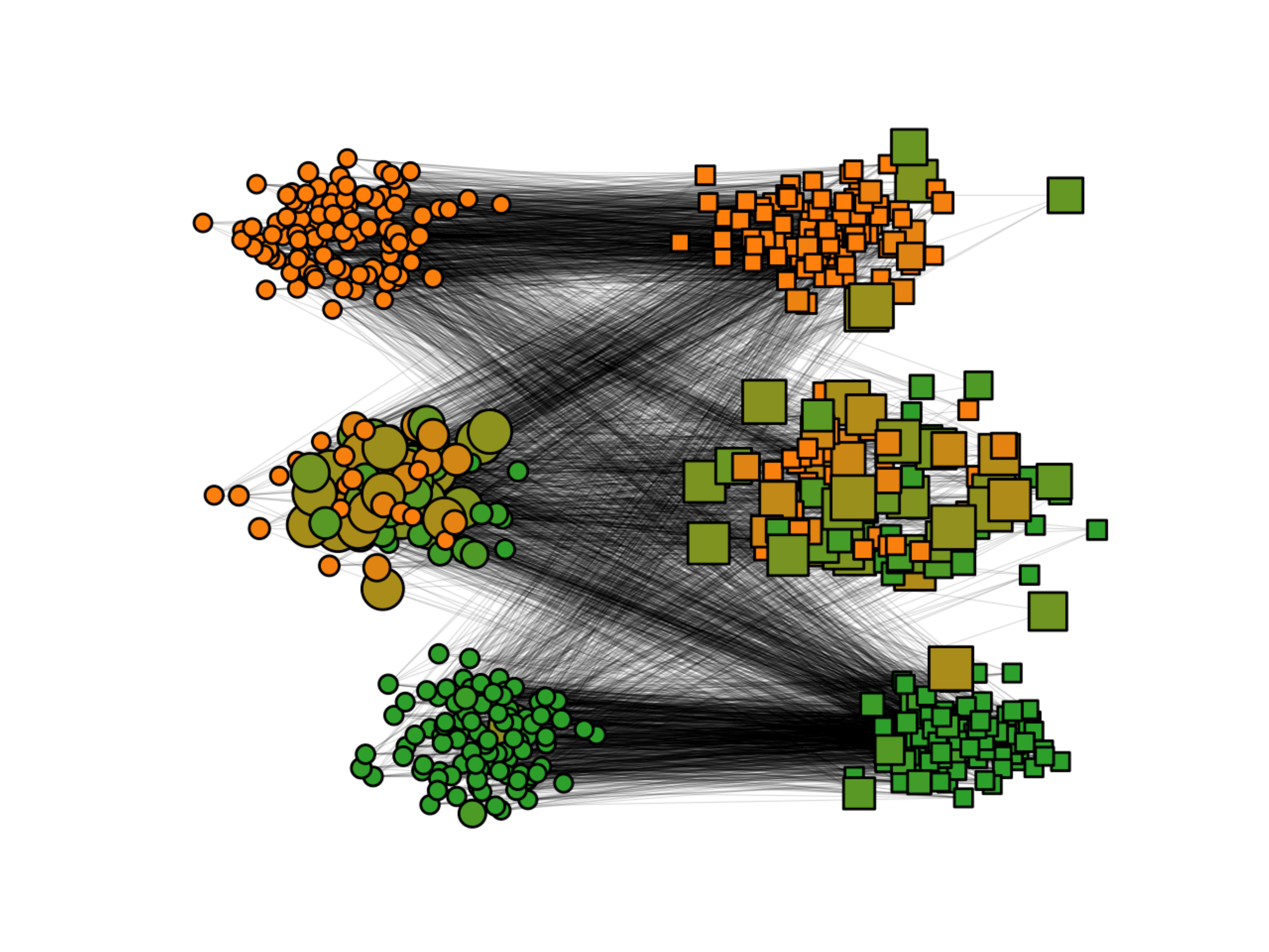} & 
\includegraphics[width=0.45\textwidth]{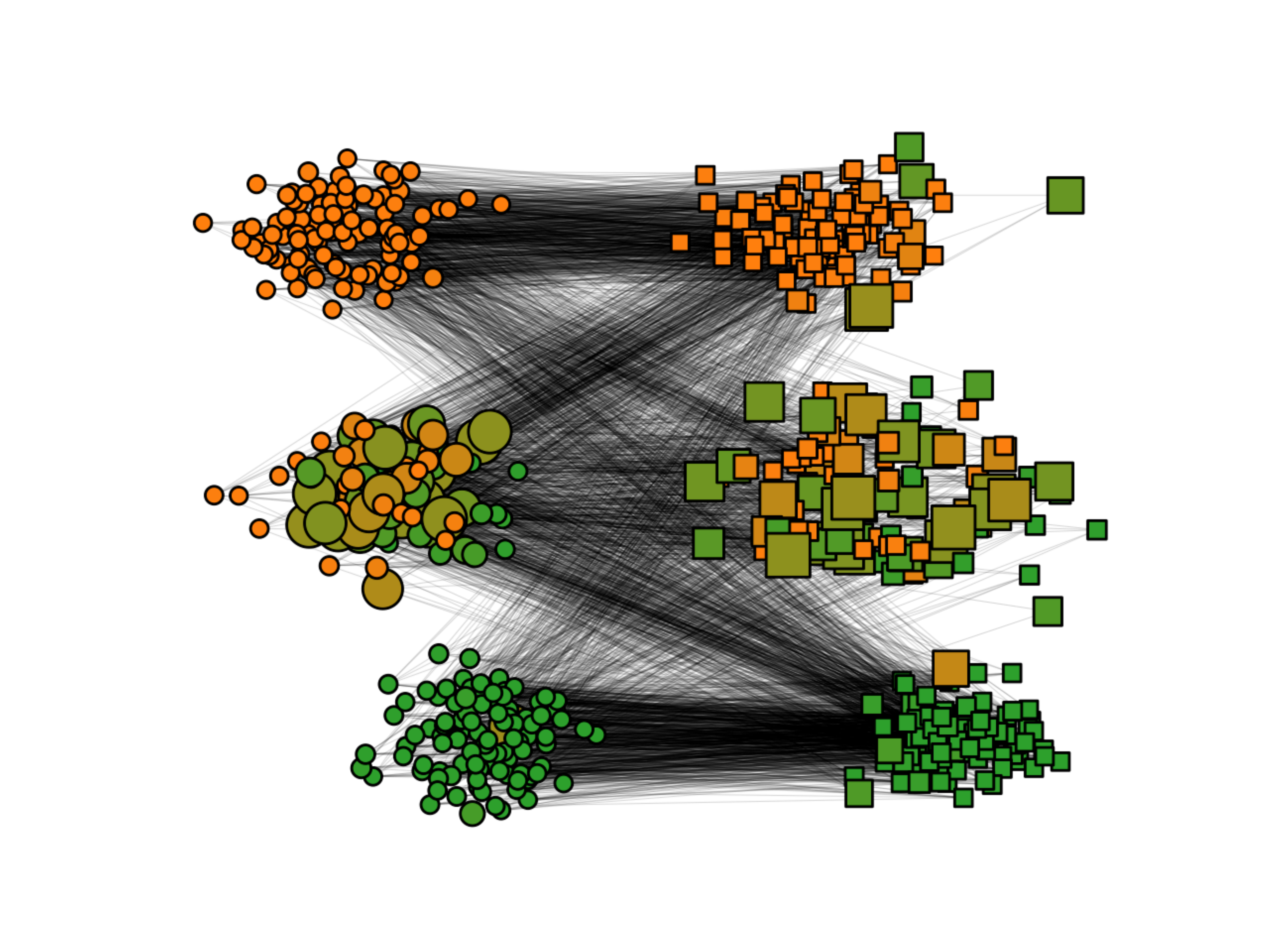} \\
(c) & (d) \\
\includegraphics[width=0.45\textwidth]{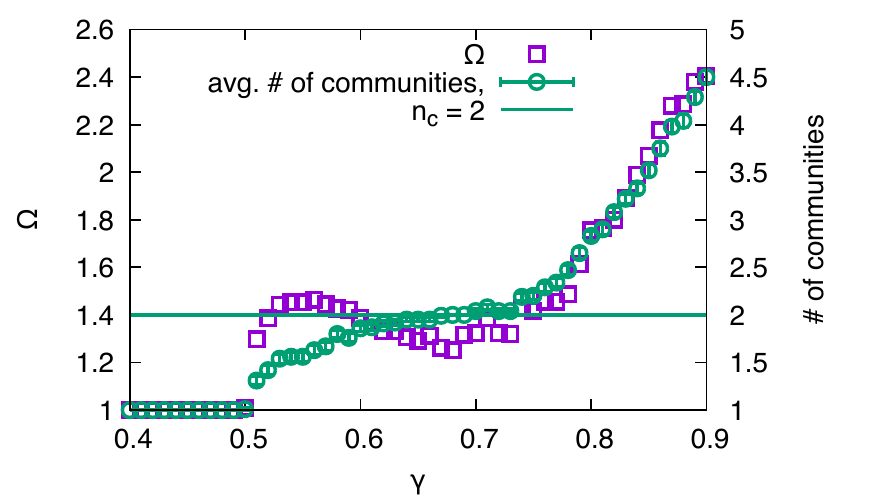} & 
\includegraphics[width=0.45\textwidth]{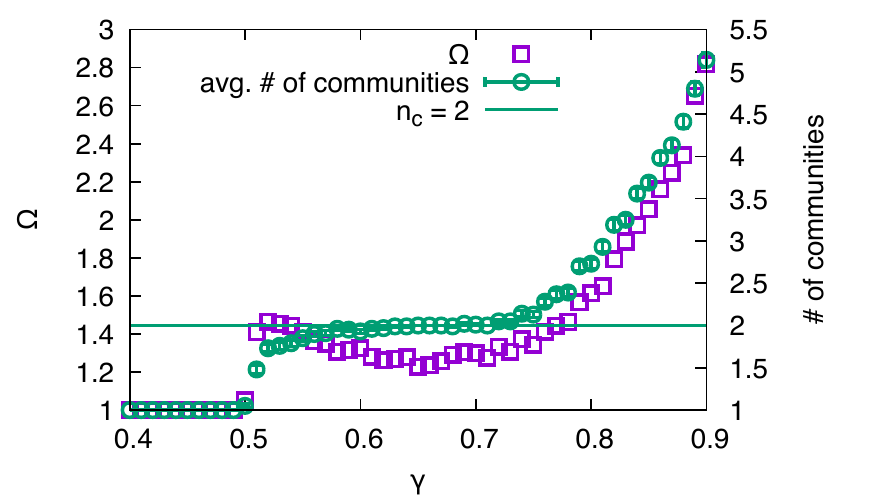} \\
\end{tabular}
\caption{Comparison between the results with different null-model terms for the 3vs3 model. Most likely community structures are shown in the case of (a) the NG null-model term\cite{Newman2004} at $\gamma = 0.68$ and (b) the BB null-model term\cite{Barber2007} at $\gamma = 0.65$, where the color inside each node represents the community where it belongs (mixed color for inconsistent communities) and its size is proportional to its local MeI value introduced in Ref.\cite{DLee2021}. We use the Fruchterman-Reingold force-directed algorithm\cite{FR} (using the ``spring layout'' function of the \texttt{NetworkX} package\cite{NetworkX}) and readjust the location to separate planted communities. The PaI curves and the average number $n_c$ of communities against the resolution parameter $\gamma$ are shown in the panels for (c) the NG null-model term\cite{Newman2004} and (d) the BB null-model term\cite{Barber2007}.} 
\label{fig:modelv}
\end{figure*}

\begin{figure*}
\begin{tabular}{ll}
(a) & (b) \\
\includegraphics[width=0.45\textwidth]{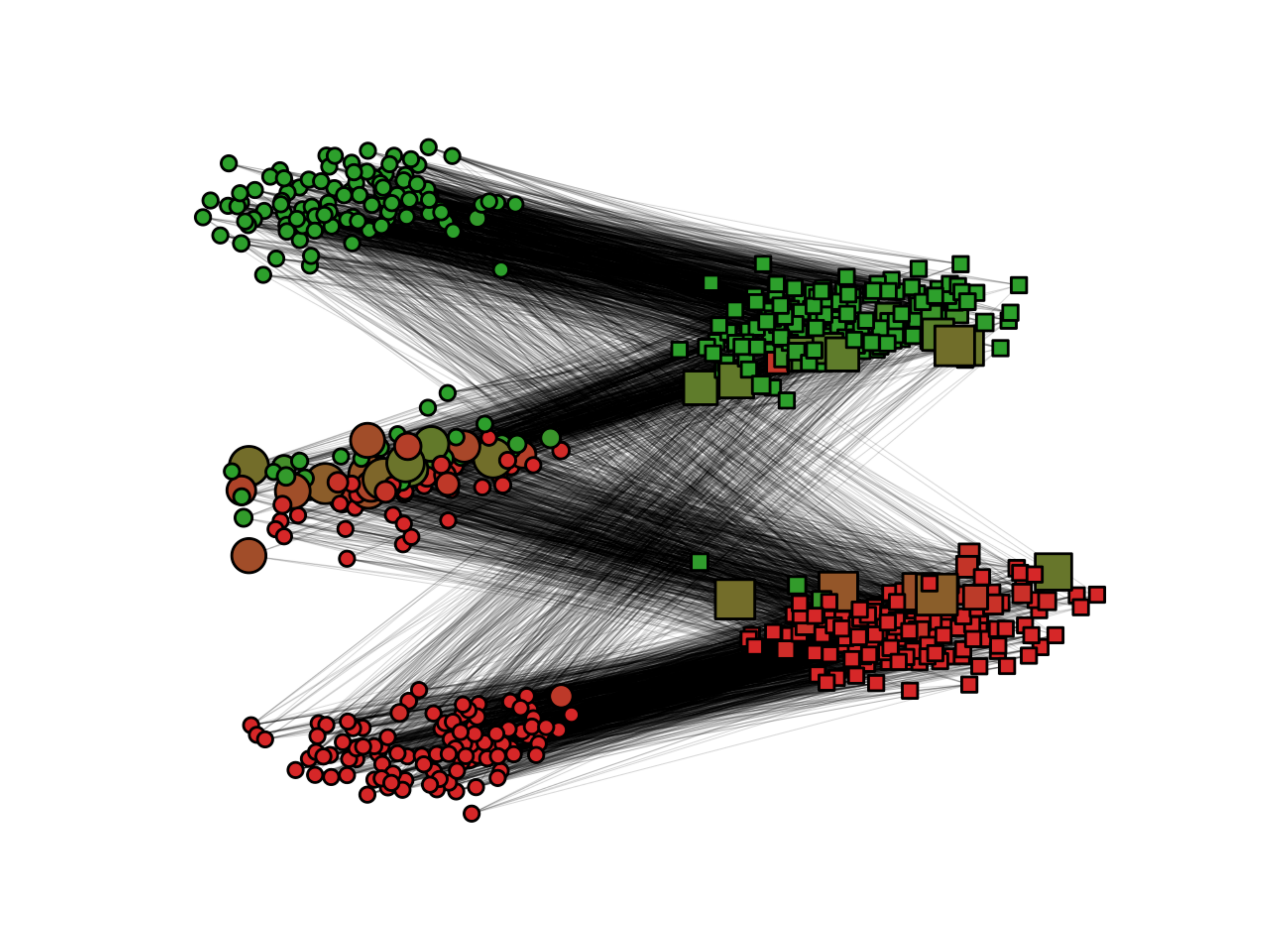} & 
\includegraphics[width=0.45\textwidth]{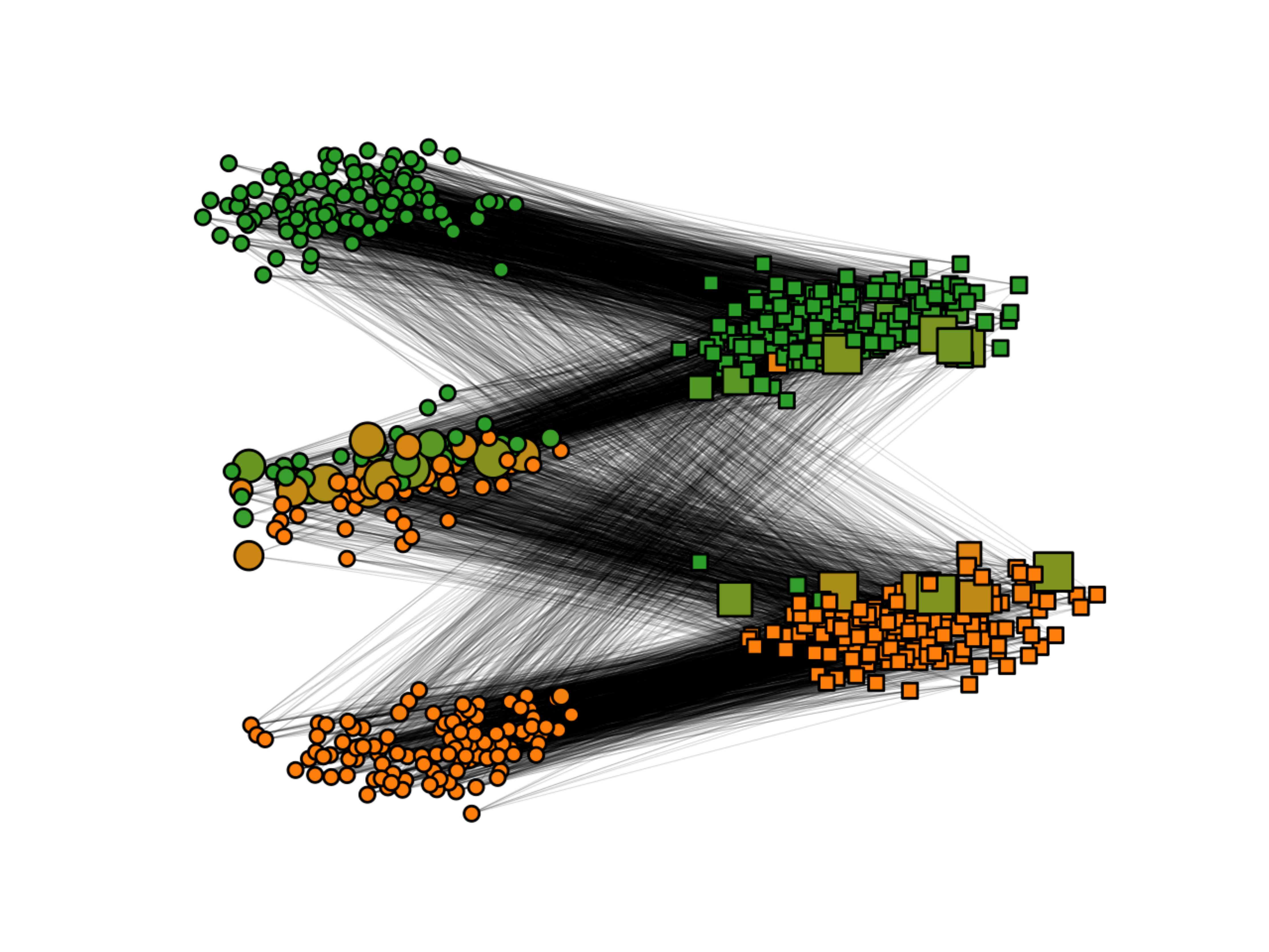} \\
(c) & (d) \\
\includegraphics[width=0.45\textwidth]{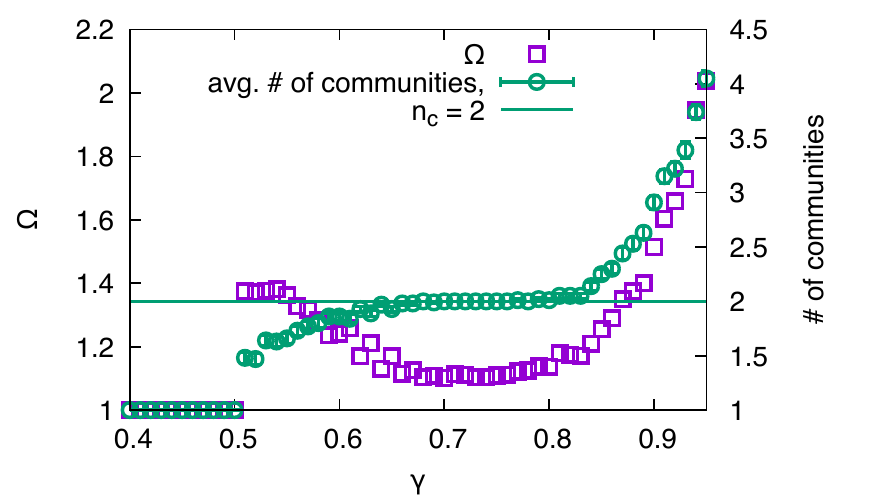} & 
\includegraphics[width=0.45\textwidth]{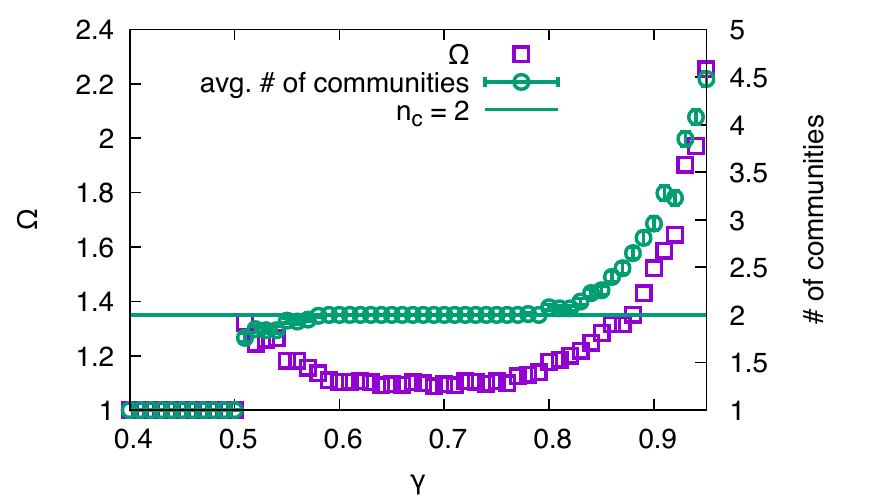} \\
\end{tabular}
\caption{The same plots as in Fig.~\ref{fig:modelv} for the 3vs2 model. Most likely community structures are shown in the case of (a) the NG null-model term\cite{Newman2004} at $\gamma = 0.73$ and (b) the BB null-model term\cite{Barber2007} at $\gamma = 0.7$. } 
\label{fig:model3v2}
\end{figure*}

\begin{figure*}
\begin{tabular}{ll}
(a) & (b) \\
\includegraphics[width=0.45\textwidth]{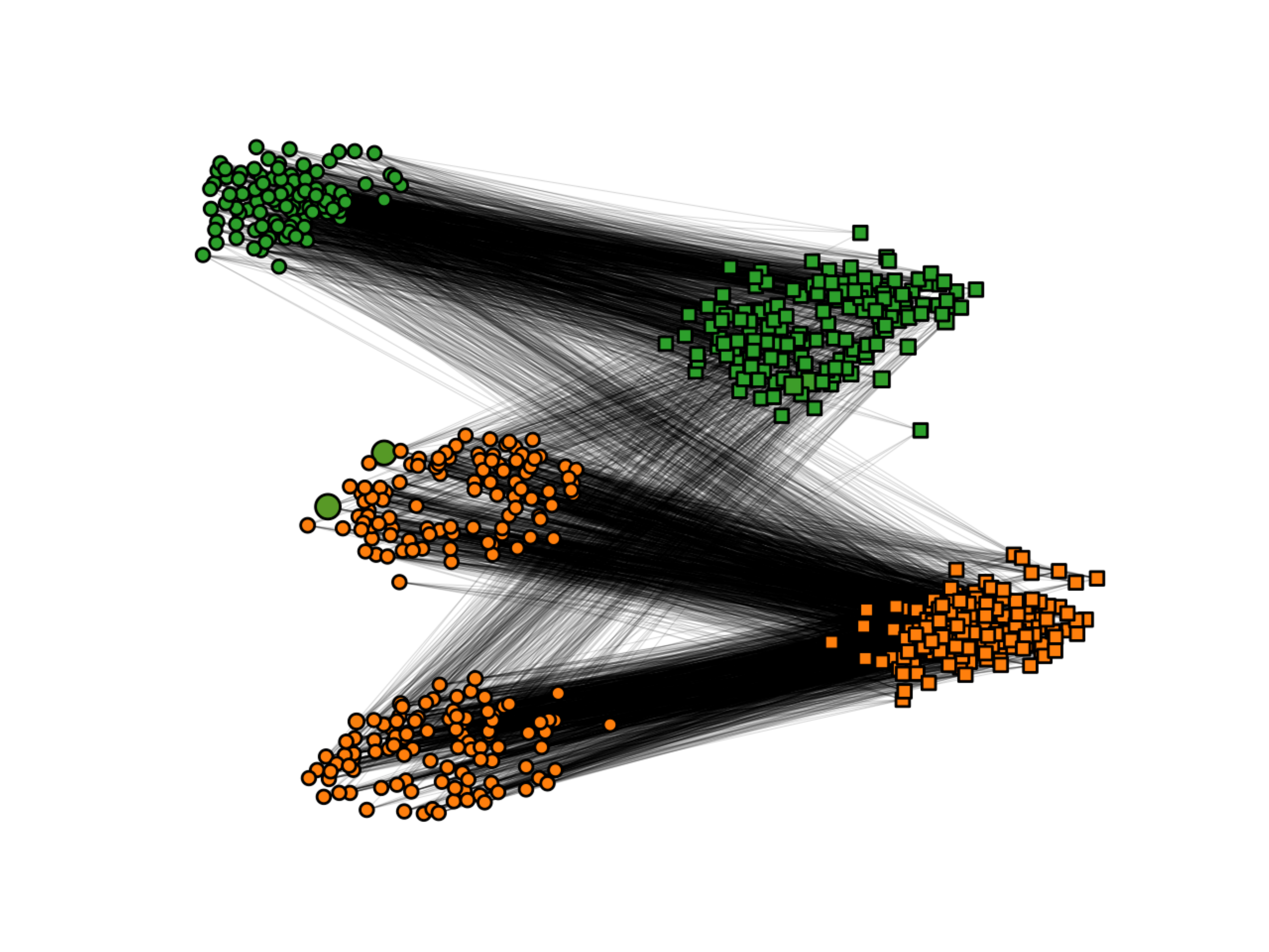} & 
\includegraphics[width=0.45\textwidth]{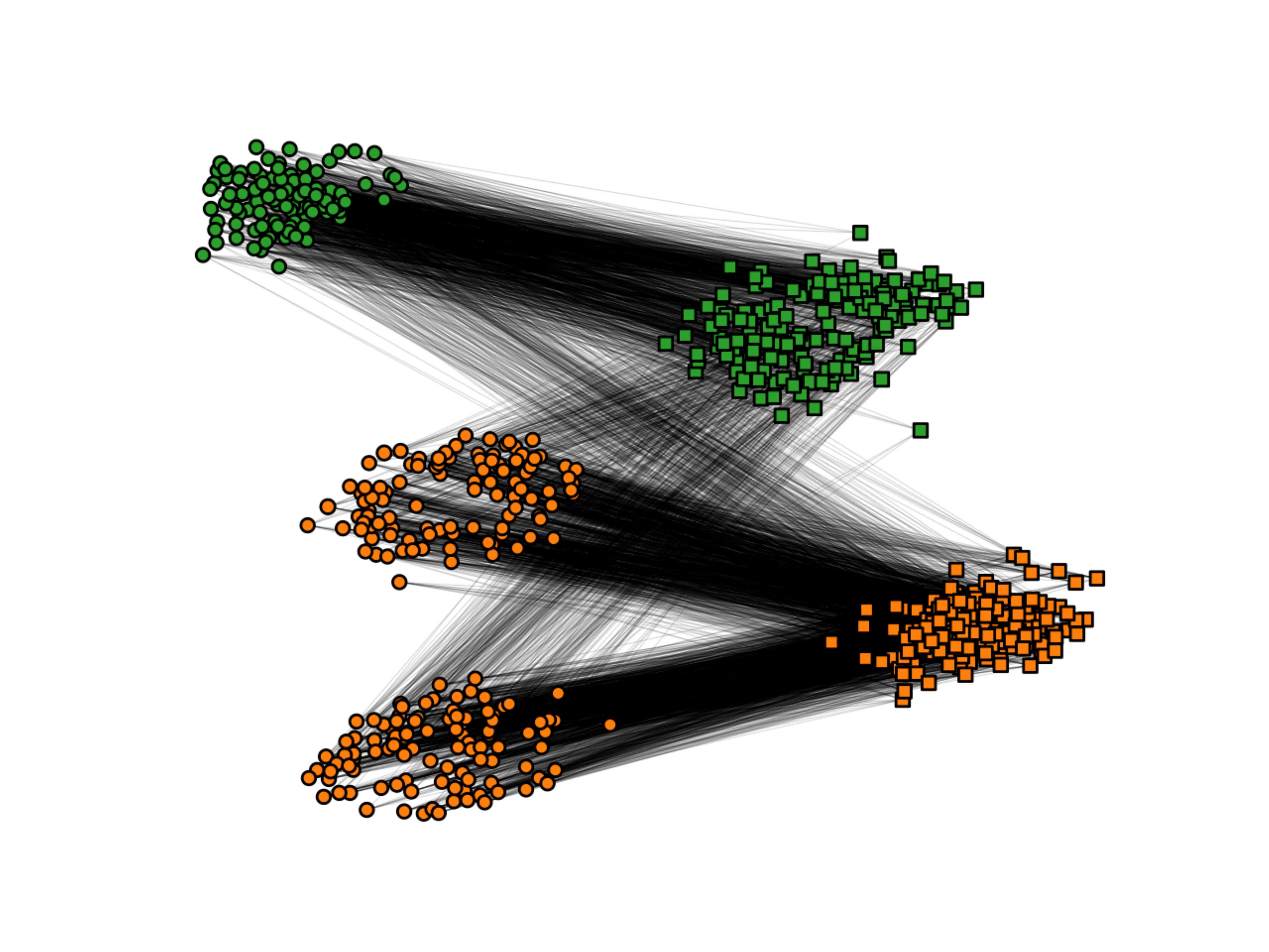} \\
(c) & (d) \\
\includegraphics[width=0.45\textwidth]{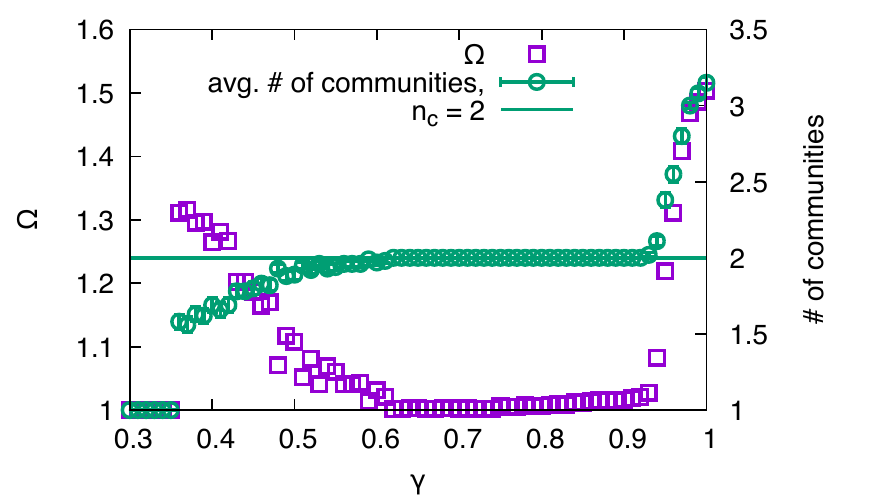} & 
\includegraphics[width=0.45\textwidth]{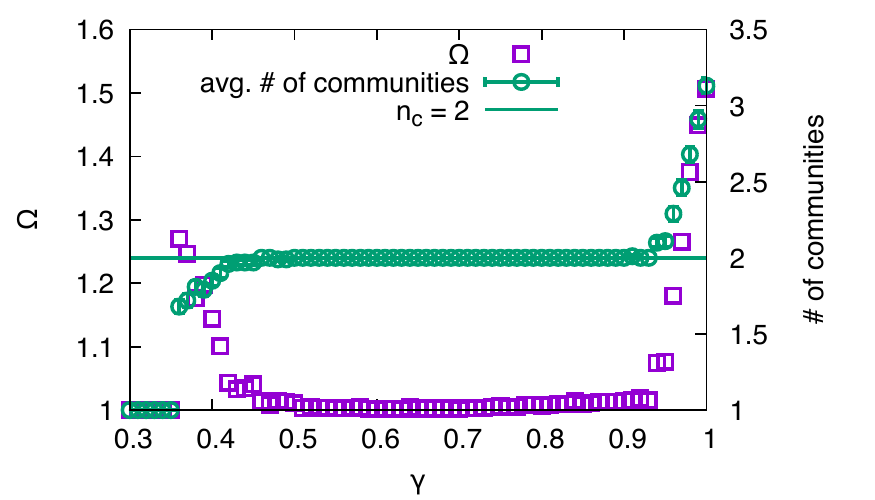} \\
\end{tabular}
\caption{The same plots as in Fig.~\ref{fig:modelv} for the 3vs2BigSmall model. Most likely community structures are shown in the case of (a) the NG null-model term\cite{Newman2004} at $\gamma = 0.75$ and (b) the BB null-model term\cite{Barber2007} at $\gamma = 0.7$.} 
\label{fig:model3v2BigSmall}
\end{figure*}

As the first test, we use a series of bipartite model networks with planted community structures, almost literally a direct realization of Fig.~\ref{fig:illustration}(a). For a bipartite network model, we initially prepare two different types of nodes that will be connected only between nodes of different types. Our first bipartite network model called the ``3vs3 model'' is composed of $600$ nodes in total, each half of which constitutes a network type. (Thus, there are $300$ nodes for each type.) Each type of node is located either on the left (circular nodes) or right (square nodes) of the network illustrated in Figs.~\ref{fig:modelv}(a) and (b). The $300$ nodes of each type are further divided into three different equal-sized groups, with $100$ nodes for each group, vertically spaced in Figs.~\ref{fig:modelv}(a) and (b). The uppermost and the lowermost groups of different types are randomly connected to each other on the same vertical level with the probability $p_\mathrm{in} = 0.1$ (the connection is only between nodes of different types of course, according to the definition of bipartite networks), and they are randomly connected to the groups located on different vertical levels with the probability $p_\mathrm{out} = 0.02$. The nodes in the middle group (in terms of the vertical level) are randomly connected to any node of a different type with the probability $p_\mathrm{mid} = 0.06$. 

With this construction, for the 3vs3 model, we have two clear-cut communities (the uppermost and lowermost ones) and one group of nodes (the middle one) without any community identity, which we intentionally put to simulate noisy community-inconsistent parts [the local inconsistency is shown by large membership inconsistency (MeI)\cite{DLee2021} values in Figs.~\ref{fig:modelv}(a) and (b)]. The comparative inconsistency curves from the NG and the BB types of null-model terms [Figs.~\ref{fig:modelv}(c) versus \ref{fig:modelv}(d)] share a common property that the inconsistency shows a local minimum In the range of $\gamma$ that yield two communities ($n_c = 2$), which is supposedly detecting the planted communities. As discussed in Ref.\cite{DLee2021}, this type of local minima of PaI in the range of the flat value (an integer value, ideally) for the number of communities is a characteristic signature of statistically reliable communities with a small inconsistency level. However, if we observe the range of the resolution parameter $\gamma$ allowing those reliable communities, the BB null-model term [Figs.~\ref{fig:modelv}(d)] produces a notably wider range of $\gamma$ for valid community detection than the NG null-model term [Figs.~\ref{fig:modelv}(c)]. 

In other words, appropriate communities can be detected with the NG null-model term (as in previous works\cite{SHLee2010,WCai2020,SHLee2020}),just as with the BB null-model term, \emph{if we carefully choose the resolution parameter}. At the same time, however, please note that if we use the BB null-model term tailor-made for treating underlying bipartite structures, the detected community structure is much less sensitive to the choice of $\gamma$, which signifies the importance of constructing the null-model term $P_{ij}$ in Eq.~\eqref{eq:Gmodularity} by using as much information as possible. The same result is observed for other variants of the model. 

The ``3vs2 model'' shown in Fig.~\ref{fig:model3v2} is composed of three groups on the left and the two groups on the right. The uppermost and the lowermost groups on the left are connected to the upper and lower groups on the right with the probability $p_\mathrm{in} = 0.1$, respectively (and the cross-connection between the upper and the lower groups occurs with the probability $p_\mathrm{out} = 0.02$). Moreover, the middle group on the left is connected to any group on the right with the probability $p_\mathrm{mid} = 0.06$. The ``3vs2BigSmall model'' shown in Fig.~\ref{fig:model3v2BigSmall} is also composed of three groups on the left and the two groups on the right, but for this model the middle group on the left is connected more densely with the lower group on the right with the probability $p_\mathrm{mid} = 0.06$ than with the upper group on the right with the probability $p_\mathrm{out} = 0.02$. The resultant 3vs2BigSmall model is composed of two clear-cut communities with different relative sizes. 

For both models (3vs2 and 3vs2BigSmall), the results in Figs.~\ref{fig:model3v2} and \ref{fig:model3v2BigSmall} support the same conclusion: both types of null-model terms (NG and BB) can produce planted communities, but the BB null-model term always allows wider ranges of $\gamma$ including reliable community structures (in these cases, the planted or the ground-truth communities). With the verification learned from the model bipartite networks, in the next subsection, we will continue the cross-examination of null-model terms by taking into consideration real networks.

\subsection{\label{sec:real_networks}Real networks}

\begin{figure*}
\begin{tabular}{ll}
(a) & (b) \\
\includegraphics[width=0.45\textwidth]{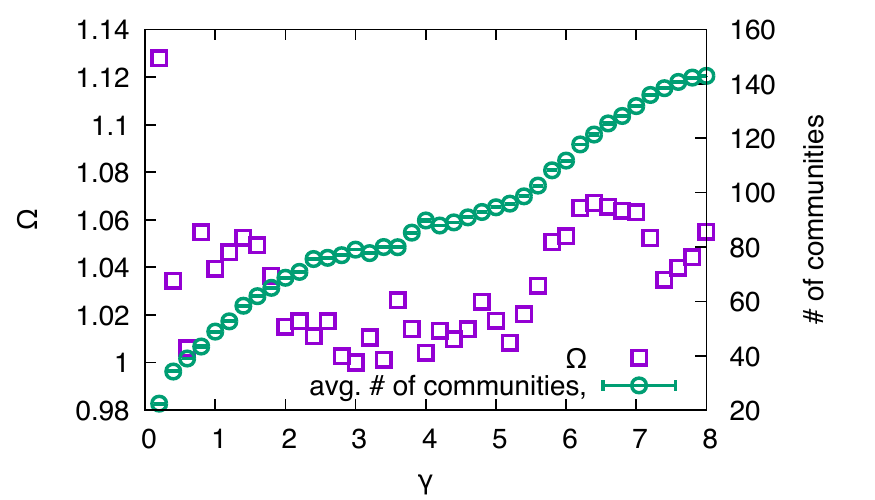} & 
\includegraphics[width=0.45\textwidth]{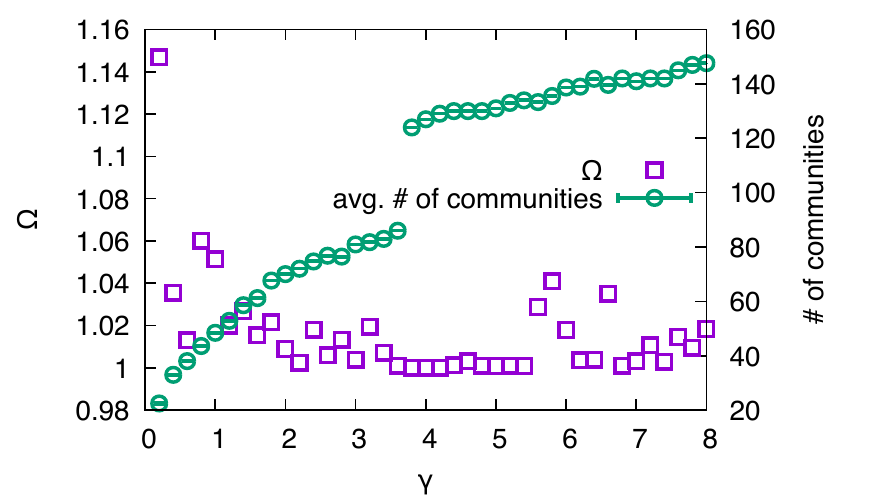} \\
\end{tabular}
\caption{Comparison between (a) the NG null-model term\cite{Newman2004} and (b) the BB null-model term\cite{Barber2007} for the network between languages and the countries in which they are spoken\cite{KONECT}.} 
\label{fig:countrylang}
\end{figure*}

\begin{figure*}
\begin{tabular}{ll}
(a) & (b) \\
\includegraphics[width=0.45\textwidth]{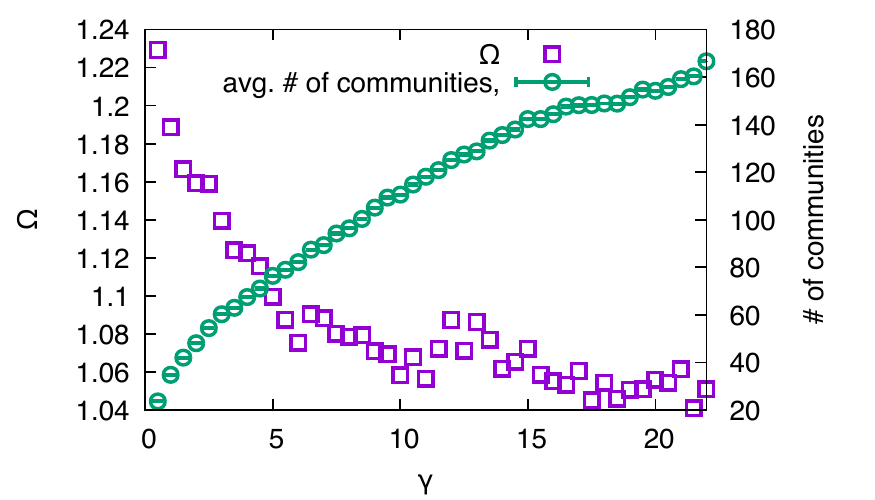} & 
\includegraphics[width=0.45\textwidth]{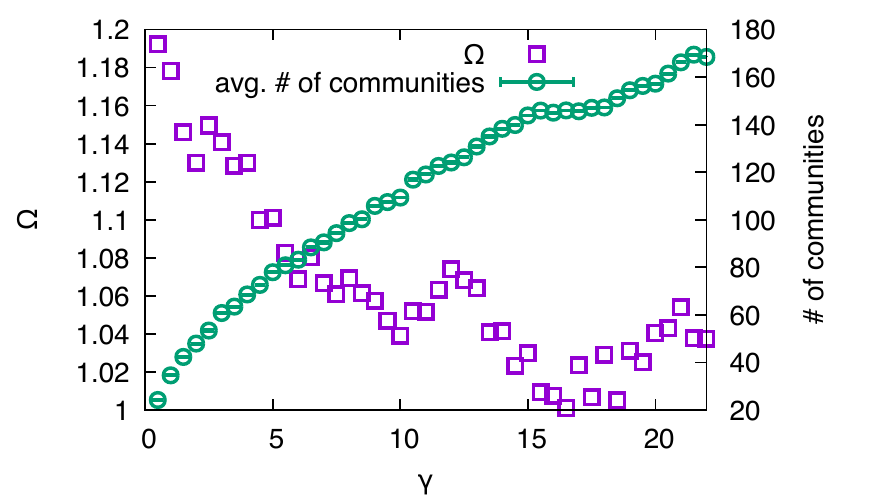} \\
\end{tabular}
\caption{Comparison between (a) the NG null-model term\cite{Newman2004} and (b) the BB null-model term\cite{Barber2007} for the network of association between human diseases and their related human genes\cite{KIGoh2007}.} 
\label{fig:diseasome}
\end{figure*}

\begin{figure*}
\begin{tabular}{ll}
(a) & (b) \\
\includegraphics[width=0.45\textwidth]{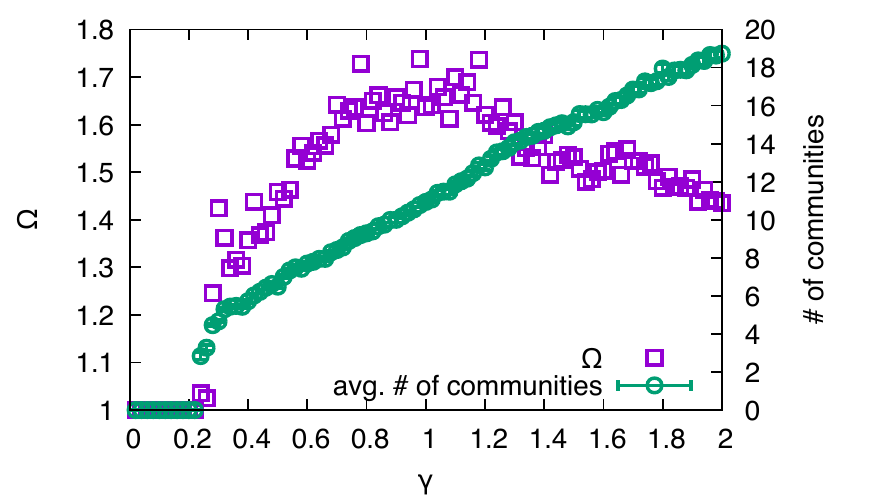} & 
\includegraphics[width=0.45\textwidth]{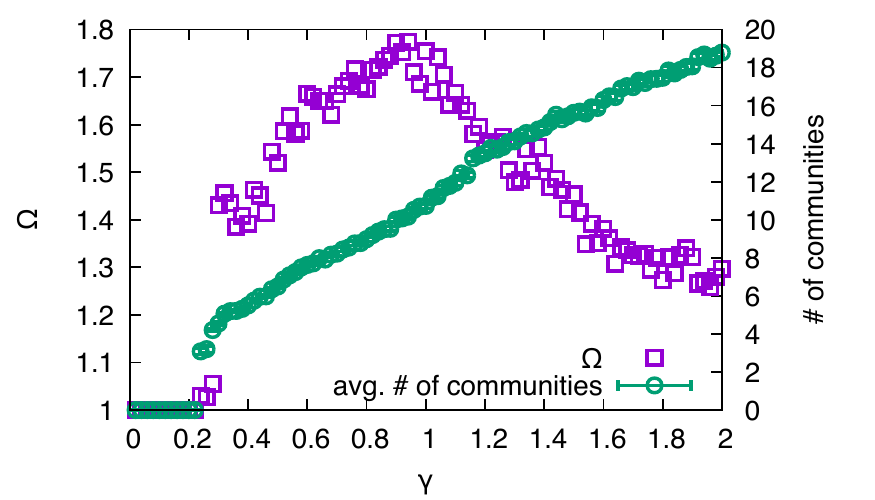} \\
\end{tabular}
\caption{Comparison between (a) the NG null-model term\cite{Newman2004} and (b) the BB null-model term\cite{Barber2007} for a mutualistic network (``M\_PL\_001'' in Ref.\cite{SHLee2020}) in ecosystems.} 
\label{fig:mutualistic}
\end{figure*}

As we have checked the importance of the proper (BB) null-model term\cite{Barber2007} for community detection in model bipartite networks, we now start to cross-check different null-model terms in cases of real bipartite networks without explicitly given ground-truth communities. The first example is the bipartite network between languages and the countries in which they are spoken\cite{KONECT}. The PaI curves and the average number of communities from the different null models in Fig.~\ref{fig:countrylang} show a similar trend to the model network results: the range of $\gamma$ yielding consistent communities (small inconsistency) is wider for the BB null-model term. In particular, we observe a sharp transition near $\gamma = 4$, as shown in Fig.~\ref{fig:countrylang}(b), i.e., jumping from $n_c \simeq 80$ to $n_c \simeq 120$, which may signal a characteristic division of linguistic cultural groups. Such a transition is not observed in the case of the NG model [Fig.~\ref{fig:countrylang}(a)], which would miss this nontrivial piece of information. Therefore, it is another example that those detailed characteristics in each system necessitate a carefully chosen null-model term for detection. 

Another example bipartite network is the association between human diseases and their related human genes\cite{KIGoh2007}, as shown in Fig.~\ref{fig:diseasome}. Although the number of communities varies with quite similar functional forms for different null-model terms, the PaI curve from the BB null-model term shows more abrupt (vertical) fluctuations (the horizontal ranges of $\gamma$ for local minima are similar in this case), which might be better at pinpointing the appropriate range of $\gamma$. 

Finally, we re-examine the community structure of bipartite mutualistic networks (pollination or seed dispersal) composed of animals and plants the author previously analyzed with the NG null-model term\cite{SHLee2020}. Based on the absence of the $\gamma$-range with a local minimum of PaI and with the flat part of the number $n_c$ of communities, the author has concluded that the community structure seems to be absent from this system, compared with other notable mesoscale structures such as core-periphery-ness\cite{Rombach2014} and nestedness\cite{Mariani2019}. As one can check from Fig.~\ref{fig:mutualistic} (a single example is presented in the figure, but we have checked all of the networks and confirmed the general conclusion), the overall functional shapes of PaI and $n_c$ against the resolution parameter $\gamma$ are qualitatively similar for both the NG and the BB null-model terms, so we conclude that the absence of a well-defined community structure in mutualistic networks can be confirmed even with the BB null-model term modified to handle bipartite networks properly. If we observe the range of large $\gamma$ values in Fig.~\ref{fig:mutualistic}(b) carefully, though, the functional shape seems to deviate systematically from the NG null-model case, so further studies might be required for a more statistically concrete conclusion.

\section{\label{sec:summary}Summary and discussion}

We have studied the efficacy of the refined null-model term inside the modularity function used for network community detection by taking a representative case of structurally constrained (but ubiquitous) bipartite networks. Despite the vast volume of literature on network community detection\cite{NewmanBook,Porter2009,Fortunato2010}, careful examinations of the null-model term in the modularity function\cite{Bassett2013} are relatively rare. Even the original work introducing the BB null-model term\cite{Barber2007} did not explicitly compare the effect of different null-model terms. In this work, we have shown the direct head-to-head comparison between the conventional NG null-model term\cite{Newman2004} and the BB null-model term\cite{Barber2007} by taking carefully constructed model bipartite networks, followed by real bipartite networks. Based on the framework to examine the self-consistency of detected network communities\cite{DLee2021}, we have shown that the BB null-model term designed to treat bipartite network communities (by encoding the aforementioned structural constraint) is, indeed, better at detecting proper communities. 

The better efficacy is especially notable in terms of the range of the resolution parameter $\gamma$ allowing reliable communities. More specifically, the range of local minima (``valleys'' in the plots) of PaI ($\Omega$) along with plateaus of the integer number of communities ($n_c$) is wider in terms of $\gamma$ for every case we examined: from the series of model networks to real networks. In other words, across different scales of communities, the BB null-mode term systematically provides literally more self-consistently reliable communities compared with the NG null-model term.

As mentioned in the first paragraph of this section, closer examinations of parts in conventional formalisms developed in network community detection or network science in general can be a profound starting point for insightful works. All of the results and speculations presented throughout the paper should be worth investigating more rigorously in terms of theoretical and empirical studies. We hope that this paper can ignite such a further series of investigations by providing an intuitive, but significant, example.

\begin{acknowledgments}
The author thanks Daekyung Lee for providing the original version of the Python codes to calculate the inconsistency measures\cite{DLee2021}. This work was supported by Gyeongsang National University Grant in 2020--2021.
\end{acknowledgments}

\end{document}